\documentclass[10pt,twocolumn,letterpaper]{article}
\pdfoutput=1
\usepackage{cvm}
\usepackage{times}
\usepackage{epsfig}
\usepackage{graphicx}
\usepackage{amsmath}
\usepackage{amssymb}
\usepackage{subfigure}

\usepackage[pagebackref=true,breaklinks=true,letterpaper=true,colorlinks,bookmarks=false]{hyperref}

 \cvmfinalcopy 

\def\cvmPaperID{259} 

 \ifcvmfinal\fi
\begin{document}

\title{Resistance-Time Co-Modulated PointNet for Temporal Super-Resolution Simulation of Blood Vessel Flows}

\author{Zhizheng Jiang, Fei Gao, Renshu Gu, Jinlan Xu, Gang Xu\\
Hangzhou Dianzi University\\
{\tt\small \{gaofei,renshugu,jlxu,gxu\}@hdu.edu.cn}
\and
Timon Rabczuk\\
Bauhaus University Weimar\\
{\tt\small  timon.rabczuk@uni-weimar.de}
}

\maketitle

\begin{abstract}
In this paper, a novel deep learning framework is proposed for temporal super-resolution simulation of blood vessel flows, in which a high-temporal-resolution time-varying blood vessel flow simulation is generated from a low-temporal-resolution  flow simulation result. In our framework, point-cloud is used to represent the complex blood vessel model, resistance-time aided PointNet model is proposed for extracting the time-space features of the time-varying flow field, and finally we can reconstruct the high-accuracy and high-resolution flow field through the Decoder module. In particular, the amplitude loss and the orientation loss of the velocity are proposed from the vector characteristics of the velocity. And the combination of these two metrics  constitutes the final loss function for network training. Several examples are given to illustrate the effective and efficiency of the proposed framework for temporal super-resolution simulation of blood vessel flows.  
\end{abstract}

\section{Introduction}

In recent years, with the advancement of medical imaging, computing power and mathematical algorithms, the rapid development of patient-specific computational fluid dynamics simulation has paved the way for a new field of computer-aided diagnosis, which provides a new framework for the prediction and prevention of atherosclerosis. Due to the complexity of blood flow biomechanics and fluid behavior, traditional CFD simulation is very time consuming.

Recently deep learning framework has been applied in many fields of computer graphics, such as generating terrains~\cite{guerin2017interactive} and high-resolution face synthesis~\cite{karras2017progressive} . In the field of  fluid simulation, machine learning or deep learning techniques also have been used to  accelerate  existing CFD solvers~\cite{ladicky2015data,tompson2017accelerating,xie2018tempogan}. However, in the previous deep learning framework for flow simulation,  the blood vessel geometry was generally converted into a depth image or voxel representation. The  image representation will lose the spatial information of the 3D  flow data, while the voxel representation requires higher computing power for more complex blood vessel models.Voxel representation will also have significant geometry error over the smooth boundary surface of the blood vessel model.  Different from previous work, point cloud  is used  as data format in our framework for the blood vessel geometry~\cite{qi2017pointnet}.

In traditional computational fluid dynamics methods,  how to capture the intricate details of blood vessel flows has been a longstanding challenge for numerical simulations and medical applications. Enormous computational costs are required to recover such details with discretized volumetric mesh models. Therefore, it is an interesting problem  to reconstruct high-resolution flow data from grossly under-resolved input data from the view of  the temporal information.  In this paper, We propose a novel deep-learning framework  for the temporal super-resolution flow simulation problem of the blood vessel model. Our main contribution can be summerized as follows:
\begin{itemize}
    \item  A novel deep learning framework is proposed for temporal super-resolution simulation of blood vessel flows with point-cloud representation. 
    \item  Resistance-time aided PointNet model is proposed for extracting the time-space features of the time-varying flow field, and the high-resolution flow field is reconstructed through the Decoder module.
    \item A training data set is constructed for the patient-specific flow field simulation on the real aorta and iliac arteries by SimVascular~\cite{lan2018re}.  
    \item A proper loss function is proposed for network training, which is a combination of the 
    the amplitude loss and the orientation loss of the velocity for blood vessel flow. 
\end{itemize}

The remainder of the paper is structured as follows. A review of related work on data-driven flow simulation and super-resolution reconstruction is presented in Section  \ref{sec:related}. Section \ref{sec:framework} presents the details of the proposed framework, including the data generation, network architecture and loss function. Several examples for super-resolution flow simulation are presented in Section \ref{sec:example}. Finally, this paper is concluded and future work is outlined in Section \ref{sec:conclude}.

\section{Related Work}

 \label{sec:related}

In this section, we will review some related work on super-resolution techniques of flow simulation.   
 
\noindent \textbf{Data-driven fluid simulation} Numerical simulation of fluids described by Navier-Stokes equations plays an important role in modeling many physical phenomena, such as dynamic medical image, climate and aerodynamics. However, solving the Navier-Stokes equations at scale remains daunting, limited by the computational cost of resolving the smallest spatio-temporal features. Recently, various methods to generate accurate results with low computational cost and efficient flow simulation methods based on machine learning and DNNs have been proposed. Ladick'y et al.~\cite{ladicky2015data} proposed a fluid simulation method based on Regression Forests and handcrafted features. Tompson et al.~\cite{tompson2017accelerating} and Xiao et al.~\cite{xiao2018adaptive}  used the DNN models to replace the pressure projection, which is an expensive computational cost simulation stage. In addition, some work has made good progress in encoding fluid simulation into simplified representations. Kim et al.~\cite{kim2019deep} proposed a generative encoder-decoder model to synthesize fluid simulations from a set of reduced parameters. Wiewel et al~\cite{wiewel2019latent, wiewel2020latent} presented an LSTM-CNN hybrids model to generate a stable and controllable temporal evolution of a fluid simulation from a latent space. Sun et al  \cite{SUN2020112732} proposed a physics-constrained deep-learning approach for surrogate modeling of fluid flows without relying on any simulation data. In their framework, a structured deep neural network (DNN) architecture is devised to enforce the initial and boundary conditions, and the governing  Navier–Stokes equations are incorporated into the loss of the DNN to drive the training. Kochkov et al \cite{Kochkove2101784118}  used end-to-end deep learning to improve approximations inside computational fluid dynamics for modeling two-dimensional turbulent flows . 
Most of previous research focused on reducing the computational cost of generating a single frame, the proposed method generates a high frame-rate flow simulation by temporal interpolation of a low frame-rate simulation. In addition, since the proposed method uses a low frame-rate flow simulation computed by a physics-based simulation, it is also possible to generate a stable high frame-rate flow simulation without cumulative errors caused by iterative DNN inference. 

\noindent \textbf{Super-reslolution framework for flow simulation}  The current super-resolution problem for flow simulation is similar with video super-resolution. The main task of video interpolation is to generate intermediate frames between the original input images, which can upsample low frame-rate videos in the time dimension to obtain higher frame-rate videos. It is an important problem in computer vision because it can overcome the time limit of the camera sensors and can be used in a variety of scenarios, such as medical imaging~\cite{karargyris2010three}, virtual reality ~\cite{anderson2016jump}, video editing~\cite{ren2018deep,zitnick2004high} and motion deblurring~\cite{brooks2019learning, xu2017motion}. In the current related work, the method of combining interpolation with deep learning is more common in the field of image processing. 
For the work of fluid simulation combined with interpolation, Raveendran et al.~\cite{raveendran2014blending} present a method for smoothly blending between existing liquid animations. Thuerey~\cite{thuerey2016interpolations} employed a five-dimensional optical flow solver to calculates a dense space-time deformation using grid-based signed-distance functions of the inputs. Sato et al. ~\cite{sato2018editing} applied interpolation to edit or combine the original fluids. The above interpolation algorithms are designed for spatial dimension, which are not suitable for solving the temporal incoherent problems. TempoGAN~\cite{xie2018tempogan} is the first method to synthesize four-dimensional physics fields with neural networks. Although TempoGAN successfully uses neural networks to solve the physical field, its goal is to obtain higher visual effects at a lower cost, and the accuracy of the simulation can be appropriately relaxed. Ferdian et al \cite{NEURIPS2018_f5f8590c} proposed a novel deep learning network to increase the spatial resolution of 4D flow MRI, trained on purely synthetic 4D flow MR data.  The corresponding deep learning network was trained to learn the mapping from noisy low resolution to noise-free HR phase images. Gao et al. \cite{Gao2021} presented a novel physics-informed DL-based spatial SR solution using convolutional neural networks, which is able to produce HR flow fields from low-resolution inputs in high-dimensional parameter space. The above-mentioned methods are all applied in the field of simulation animation, but for hemodynamics, more attention should be paid to the accuracy of simulation. In this article, we use the Resistance-Time Co-modulation (RTCM) module to integrate the simulation boundary conditions and time frame information into global variables, so that our network can better express the physical characteristics of the flow field.

\noindent \textbf{Deep learning with point cloud representation} In recent years, several famous deep learning architectures, such as Pointnet~\cite{qi2017pointnet}, Pointnet++~\cite{qi2017pointnet++} and PointCNN \cite{NEURIPS2018_f5f8590c}, have proposed for  the intelligent processing of unordered point sets, and there are more and more point cloud-based work including 3D shape completion~\cite{huang2020pf} and segmentation~\cite{xu2020sceneencoder}. Aiming at the unordered of point clouds, loss functions such as Chamfer Distance(CD)~\cite{fan2017point} and Earth Mover’s Distance(EMD)~\cite{fan2017point} are proposed.Most of the previous research aimed at 3D flow field simulation on regular voxel grids. However, due to the data sparsity and computational cost of 3D convolution, the voxel-based method is limited by its resolution. The proposed method uses a point cloud form to represent the vascular flow field, which can make better use of the limited computational cost and complete high-precision simulation work.   

\section{Proposed Method}
\label{sec:framework}
\subsection{Data Generation}
\begin{figure}
    \centering
    \includegraphics[scale=.45]{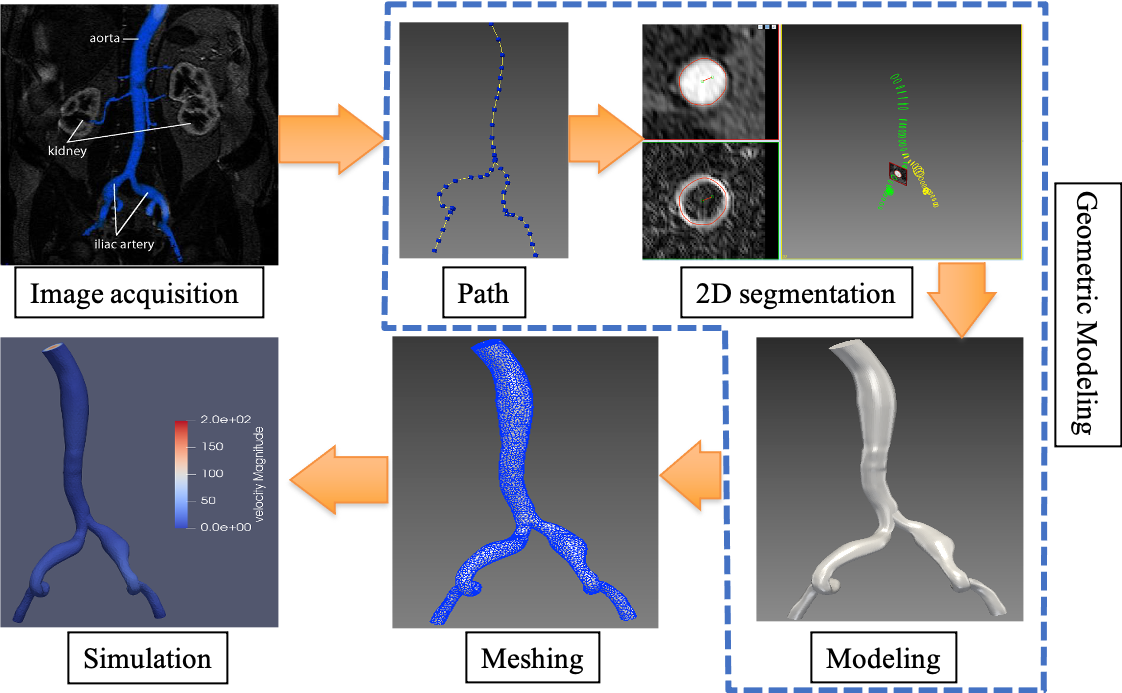}
    \caption{Simulation flowchart}
    \label{fig:sim_flow}
\end{figure}

In our method, time-varying flow field can be constructed via image acquisition, modeling, meshing, and simulation, where SimVascular~\cite{lan2018re} is used for blood flow simulation. Figure \ref{fig:sim_flow} shows the simulation procedure.

\textbf{Image acquisition.}
The MRI images of the aorta and iliac artery provided by official website of SimVascular~\cite{lan2018re} are used in our framework, and the aorta and iliac artery blood vessels are extracted from these MRI images.

\textbf{Modeling.}
In order to model the blood vessel, skeletons along the approximate centre-lines of aorta and iliac artery of the given MRI image are created first. Then, a set of vessel contours generated through segmentation of imaging data are formed along the path of skeleton. These contours are used to create surface of 3D blood vessel model. Moreover, different blood vessel models can be created by modifying the contours.

\textbf{Meshing.}
The generated model through modeling is boundary representation, hence TetGen~\cite{si2015tetgen} software is used to perform meshing operations to create 3D representation. In the process of meshing, the final number of grid nodes generated for each blood vessel is almost the same by adjusting the Global-Max-Edge-Size parameter, which guarantees the consistency of the input size of the network. The parameters of the grid generation for each blood vessel model are shown in Table \ref{tab:mesh}.

\begin{table}
\begin{center}
\begin{tabular}{|l|c|c|c|c|c|c|}
\hline
Model & Aorta1 & Aorta2 & Aorta3 & Aorta4 & Aorta5 \\
\hline\hline
GMES & 0.377 & 0.312 & 0.350 & 0.412 & 0.308 \\
Nodes & 8197 & 8204 & 8207 & 8198 & 8192 \\
Elems & 38098 & 36543 & 37187 & 38258 & 36259\\
Edges & 11850 & 12996 & 12519 & 11796 & 13215\\
Faces & 7900 & 8664 & 8346 & 7864 & 8810\\
\hline
\end{tabular}
\end{center}
\caption{Meshes generated by TetGen. GMES: Global-Max-Edge-Size; Nodes: number of mesh nodes; Elems: mesh elements; Edges: number of mesh edges; Faces: number of mesh faces}
\label{tab:mesh}
\end{table}
\textbf{Simulation.}
In our work, we use incompressible Navier-Stokes equation to simulate the blood flow:
\begin{equation}
    \begin{array}{l}
        \rho \dot{v}_{i}+\rho v_{j} v_{i, j}-p_{, i}-\tau_{i j, j}=0   \\
        v_{i, j}=0 \\ 
    \end{array},
    \label{eq:ns}
\end{equation}
Where $\rho$ is the blood density,$v_{i}$ is $i$-th component of the velocity field and $\dot{v}_{i}$ is its time derivative, $\mathrm{p}$ is the pressure, and $\tau_{i, j}$ is the viscous part of the stress tensor.

In this work, resistance boundary condition~\cite{vignon2006outflow} is set in the simulation process. For the same blood vessel model, different time-varying flow fields are generated by modifying parameters of resistance boundary conditions. By setting different time steps, simulation results with different accuracy are obtained. In order to satisfy convergence, time step is set to 0.001s and generate 1000 frames of  time-varying flow field, where 250 frames are extracted as low-accuracy data. For high-accuracy time-varying flow field, time step is set to 0.0001s, and 500 frames are extracted from generated 10,000 frames.

\textbf{Point cloud extraction.}
The flow field generated by SimVascular simulation is represented on tetrahedral mesh, we hope to express the flow field in point cloud. Therefore, vertices of tetrahedral mesh which include vertex coordinates and flow field information such as velocity field, are used as point cloud input in the following network. In addition, number of points in each point cloud should be the same for the sake of consistency of the input data with the data set. In order to better describe the time-varying flow field and the geometric details of blood vessels, we set the number of points in each point cloud to 8192. If the nodes of blood vessel generated by meshing is more than 8192, nodes with sequence number greater than 8192 can be deleted directly. Since we tried to control the number of vertices close to 8192 in the mesh generation, the deletion operation will not have great impact on the flow field data.

\textbf{Total amount of data.}
In the flowchart of simulation, 5 different blood vessel models are obtained by modifying the contours in the 2D segmentation step. For each blood vessel model, 20 different resistance boundary conditions and 2 different time steps (0.001s, 0.0001s) are used to construct the flow fields, where the generated time-varying flow fields with time step 0.001s are low-accuracy data, and flow fields with time step 0.0001s are high-accuracy data. 
In the interpolation experiment, there are $5 \times 20 \times 250=25000$ frames are generated as low-accuracy data and $5 \times 20 \times 500=50000$ frames are created as high-accuracy data. For the velocity field of each blood vessel at the same time, the low-accuracy velocity field is used as the network input, and the high-accuracy velocity field is used as the ground-truth.

\begin{figure*}
    \centering
    \includegraphics[scale=0.5]{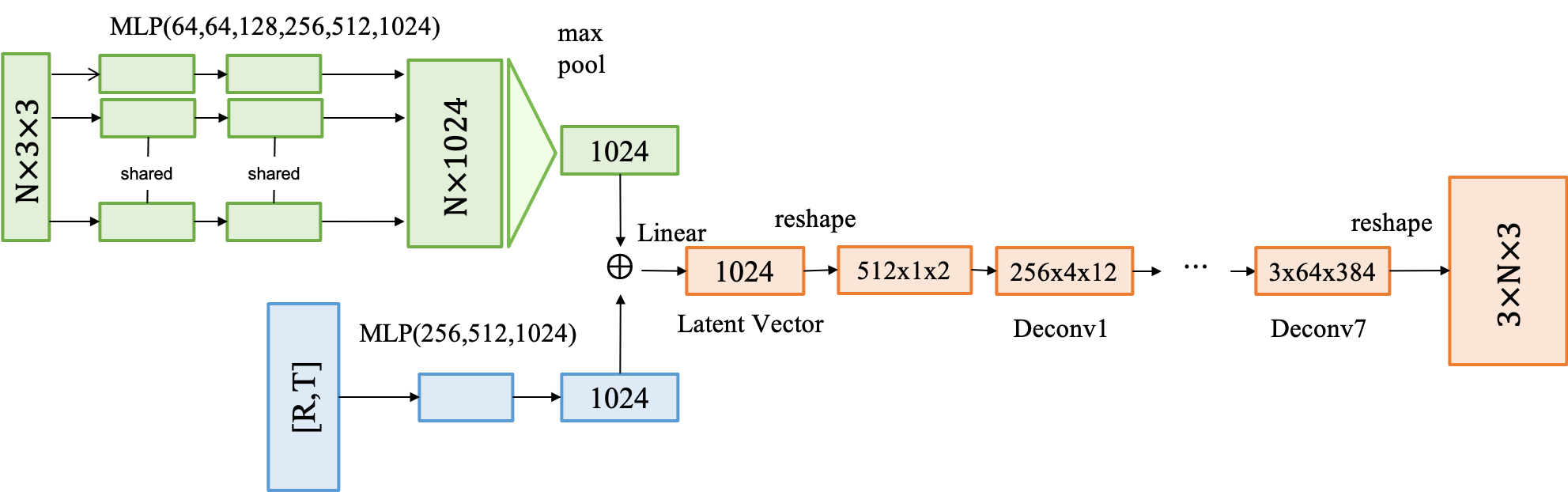}
    \caption{Architecture of resistance-time co-modulated PointNet for temporal super-resolution simulation.}
    \label{fig:netarc}
\end{figure*}

\subsection{Network Architecture}
\label{ssec:netarc}

\textbf{Problem Statement.} 
The input of our network is a velocity field with low accuracy and low time-resolution, which is simulated with a large time step (as introduced in the previous section). The output of the network is the estimated velocity field with high accuracy and high time-resolution. In the implementation, we use the velocity field simulated with a small time step under the same simulation conditions as the ground-truth. Our goal is to improve the time resolution as well as the simulation accuracy of velocity fields.

For facility of description, we take the one-frame interpolation task as an example to introduce our algorithm. Note that our algorithm can solve $k$-frame interpolation task, where $k$ denotes we aim to interpolate $k$ frames between two adjacent input frames, with $k \in \mathbb{R}$. 
In the one-frame interpolation task, the input of our network includes the low-resolution velocity field, coordinates, resistance, and serial number of adjacent two frames, i.e. at time $t$ and $t+1$. The target outputs of our network are high-resolution velocity fields at time $t$, $t+0.5$, and $t+1$, sequentially. 


\textbf{Network Overview.} 
The overall network adopts Encoder-Decoder architecture, as shown in Figure \ref{fig:netarc}. Our network includes three modules: one velocity encoder $E_v$, one resistance-time encoder $E_{rt}$, and one decoder $D$. Specially, $E_v$ encodes low-resolution velocity fields to a deep feature vector $\mathbf{f}_v$. $E_{rt}$ maps the resistance and temporal information into a deep feature vector $\mathbf{f}_{rt}$. Finally, $D$ takes a concatenation of $\mathbf{f}_v$ and $\mathbf{f}_{rt}$ as input, and outputs the estimated high-resolution velocity fields. 

In the task of video interpolation, the network is typically composed of several 2D convolutional layers. However, in our task, the blood vessel flows are formulated as point clouds, which follows irregular representation structures. We cannot use 2D convolutional operators here. Traditionally, when dealing with point clouds, researchers often convert them into more regular formats such as depth images or voxels, to facilitate the definition of convolution operations for weight sharing. The proposal of PointNet \cite{qi2017pointnet} allows us to directly input point cloud data for processing. Thus the architectures of our encoders and decoder mainly follow PointNet with essential modifications. 

\textbf{Velocity Encoder.} 
First, we use a Velocity Encoder, $E_v$, to extract global features from the input point cloud. Following PointNet \cite{qi2017pointnet}, we design $E_v$ as a multi-layer perceptron (MLP) to encode the input velocity fields. Specially, we use six fully-connected layers, and map each sampling point into a 1,024-dimension feature vector. Note that all points share the same MLP parameters. Finally, all the feature vectors of the input $N$ samples are pooled to a single vector, i.e. $\mathbf{f}_{v} \in \mathbb{R}^{1,024}$, by a globally max-pooling (GMP) layer.  This feature vector represents the global information of the whole input velocity fields. 

The input of this velocity encoder is $\left[\mathbf{u}_{i, t}, \mathbf{u}_{i, t+1}, \mathbf{l}_{i}\right] \in \mathbb{R}^{3 \times 3}$, where $\mathbf{u}_{i, t}=\left[u_{i, t, x}, u_{i, t, y}, u_{i, t, z}\right]^{T}$ represents the low-resolution velocity field of the $i$-th sample at time $t$; $x$, $y$, and $z$ denote velocity components in the corresponding orientations, respectively;  $\mathbf{l}_{i}=\left[l_{i, x}, l_{i, y}, l_{i, z}\right]^{T}$ denotes the coordinates of the $i$-th sample in a point cloud. Assume that there are $N$ sampling points in a blood vessel, the dimension of the input is $N \times 3 \times 3$. 
The calculation of resistance-time feature is then formulated as:
\begin{equation}
    \mathbf{f}_{v} = \mathrm{GMP}( \mathrm{MLP}(\left[\mathbf{u}_{i, t}, \mathbf{u}_{i, t+1}, \mathbf{l}_{i}\right])).
    \label{eq:f_v}
\end{equation}

\textbf{Resistance-Time Encoder.}
\label{ssec:rscm}
Since blood flows are time-varying and highly correlated with the resistance of blood vessels, it is necessary to consider these two factors in the simulation of velocity fields. To this end, we additionally input the resistance and time information, and use an auxiliary encoder to transfer them to a deep feature vector $\mathbf{f}_{rt}$.

Let $\left[r_i, t, t+0.5, t+1 \right]$ denote resistances and time of input frames, where $r_{i}$ represents the resistance value used in the simulation of the $i$-th sample, and $t, t+0.5, t+1$ represents the serial number of flow frames. We use a 3-layer MLP as the resistance-time encoder $E_{rt}$, which maps the input vector to a 1,024-dimensional feature vector $\mathbf{f}_{rt}$.  The calculation of resistance-time feature is formulated as:
\begin{equation}
    \mathbf{f}_{rt} = \mathrm{MLP}(\left[r_i, t, t+0.5, t+1 \right]).
    \label{eq:f_rt}
\end{equation}
Afterwards, $\mathbf{f}_{rt}$ is concatenated with the global feature $\mathbf{f}_{v}$ (Eq.\ref{eq:f_v}), and input into a following decoder for prediction.

In summary, we use three types of information as inputs, i.e. the velocity fields, the resistance and time, and locations of cloud points. By integrating the velocity feature $\mathbf{f}_v$ and the resistance-time feature $\mathbf{f}_{rt}$, our model can capture the time-varying characteristics of blood vessel flows. In addition, the inputting coordinate information makes our model capable of capturing local structures from nearby points. As a result, the proposed two encoders learn spatial and temporal representations 
of low-resolution blood vessel flows, which will be fed into a following decoder for estimating high-resolution flows.

\textbf{Decoder.} 
Our decoder is mainly responsible for generating the velocity field with high accuracy and high time resolution.
The network architecture of the decoder directly affects the quality of final output velocity fields. The simplest and most intuitive method is using several fully-connected layers to gradually upsample the encoding features to a velocity field. This method is commonly used in tasks such as classification and semantic segmentation. However, such methods cannot make full use of the geometric information of the point cloud. In this work, our decoder is composed of 7 deconvolution layers, each is followed by a ReLU activation layer. In this way, the encoding features are gradually upsampled and finally mapped to a velocity field. 

The predicted flow field is expressed as:
\begin{equation}
    \hat{y}_{i} = \left[\hat{\mathbf{v}}_{i, t}, \hat{\mathbf{v}}_{i, t+0.5}, \hat{\mathbf{v}}_{i, t+1}\right] = D(\mathbf{f}_{v} \oplus \mathbf{f}_{rt}),
    \label{eq:decoder}
\end{equation}
where $\hat{y}_i \in \mathbb{R}^{3 \times 3}$, $\hat{\mathbf{v}}_{i, t}$ denotes the estimated velocity of the $i$-th point at time $t$,  $\mathbf{f}_{v}$ and $\mathbf{f}_{rt}$ are the velocity and resistance-time feature vectors, and $\oplus$ denotes the concatenation operation. The estimate velocity field is correspondingly denoted by $\hat{y} \in \mathbb{R}^{N \times 3 \times 3}$, with $N$ being the number of points.

\subsection{Loss Functions}
\label{ssec:loss}

In image super-resolution reconstruction tasks, researchers have proposed various loss functions, such as $L_{p}$ norms, multi-scale structural similarity (MS-SSIM) \cite{zhao2016loss}, and perceptual loss \cite{johnson2016perceptual}, etc. In contrast, the super-resolution reconstruction of velocity fields has not been studied yet. The velocity field is typically represented in format of directional vectors. To measure the difference between two directional vectors, it is necessary to consider both the magnitude and orientation. Therefore, we use an magnitude loss and an orientation loss to measure the difference between the estimated high-resolution velocity field $\hat{y}$ and the ground-truth $y$, and add them together as the total objective. Here, $y = \{y_i\}_{i=1}^N \in \mathbb{R}^{N \times 3 \times 3}$, where $y_{i}=\left[\mathbf{v}_{i, t}, \mathbf{v}_{i, t+0.5}, \mathbf{v}_{i, t+1}\right] \in \mathbb{R}^{3 \times 3}$ denotes the high-resolution velocity field of the $i$-th samping point at time $t$, $t+0.5$ and $t+1$, respectively. 

\textbf{Magnitude Loss.}
First, we use the velocity modulus length to represent the velocity, and use the Euclidean distance of the modulus length between $\hat{y}$ and $y$ as the magnitude loss. The magnitude loss is formulated as:
\begin{equation}
L_{mag}=\frac{1}{N} \sum_{i=1}^{N} \Big\| \| y_{i}\|_{2} - \left\|\hat{y}_{i}\right\|_{2} \Big\|_2,
\label{eq:loss_mag}
\end{equation}
where $N$ is the number of points. The magnitude loss will enforce the estimated velocity vectors having the same magnitudes as the target velocity fields. In other words, at each sampling point in the blood vessel, the estimated blood speed should be the same as that simulated with a fine time interval.

\textbf{Orientation Loss.}
We calculate the cosine distance between the estimated velocity vector and the corresponding ground-truth vector at each point. The average cosine distance over all the $N$ sampling points is adopted as the orientation loss $L_{ori}$, which is formulated as:

\begin{equation}
    L_{o r i}=\frac{1}{N} \sum_{i=1}^{N}\left(1-\frac{y_{i}^{T} \hat{y}_{i}}{\left\|y_{i}\right\| \cdot\left\|\hat{y}_{i}\right\|}\right).
    \label{eq:loss_ori}
\end{equation}

Our argument is that the blood, with respect to the estimated velocity field, should flows at the same directions as that simulated in high-resolution settings, at every sampling points. 

\textbf{Full Objective.}
We use an integration of the magnitude loss and the orientation as our full loss function:
\begin{equation}
    L_{mo} = \alpha L_{mag}+\beta L_{ori},
    \label{eq:loss_all}
\end{equation}
where $\alpha$ and $\beta$ are weighting factors. In subsequent experiments, we set $\alpha$ to 0.05 and $\beta$ to 1 as default. We optimize our network by minimizing the total loss.

\section{Experimental results}
\label{sec:example}
\subsection{Settings and performance indicators}
In the experiments, Adam optimizer is used in the training process and the learning rate is set to 0.0003. The batch size is set to 32. For the experiment of one-frame interpolation with network, the number of iterations is about 63,000, and for the experiment of two-frame interpolation, the number of iterations is about 42,000. The network uses interval adjustment learning rate (Step LR), where step size is set to 32 and gamma is set to 0.2. A random division method is adopted for the data set, and the division ratio is 8:1:1.

We use network to generate flow field with high resolution, and use linear interpolation to generate the same flow field from low resolution, then the results of the two are compared. The linear interpolation method is as follows:
\begin{equation}
\mathbf{v}_{i, c}=\frac{e-c}{e-s} \mathbf{v}_{i, s}+\frac{i-s}{e-s} \mathbf{v}_{i, e},
\end{equation}
where $c$ represents the current time, $s$ is the start time, $e$ is the end time, and these three parameters satisfy $s<$c$<e$. $\mathbf{v}_{i, s},\mathbf{v}_{i, e}$ and $\mathbf{v}_{i,c}$ are velocity field of the $i$-th sample data at time $s,e$, and $c$ respectively.

The results will be assessed from four performance indicators which are visualization results, velocity interval range, average modulus length error, and relative error. For the range of the velocity field, it includes the range of the velocity modulus length and the range of the diversion in the $x$, $y$, and $z$ directions.

The average modulus length error refers to the average modulus length difference between the generated result and the ground-truth at a certain moment. We denote it as \emph{MME}, and use this error to judge the quality of the generated result at each frame. The definition of \emph{MME} is as follows:
\begin{equation}
MME=\frac{1}{N} \sum_{k=1}^{N}\left|\left\|\mathbf{v}_{k}\right\|-\left\|\hat{\mathbf{v}}_{k}\right\|\right|,
\end{equation}
where $N$ is the number of points, $\mathbf{v}_{k}$ is the generated velocity of the $k$-th point in the current flow field, and $\hat{\mathbf{v}}_{k}$ is the ground-truth velocity of the $k$-th point in the current field.

Finally, we also use the relative error to evaluate the time-varying flow field. The relative error is expressed by \emph{RE}, and its calculation is as follows:
\begin{equation}
\begin{split}
    &RE=\frac{1}{N \times T} \sum_{t=1}^{T} \sum_{k=1}^{N} \frac{\mid\left\|\mathbf{v}_{k, t}\right\|-\left\|\hat{\mathbf{v}}_{k, t}\right\| \mid}{\left\|\hat{\mathbf{v}}_{k, t}\right\|}, \\
 &\left\|\mathbf{v}_{k, t}\right\|>10^{-4}
\end{split}
\end{equation}
where $N$ is the number of points, $T$ is the frame number of flow field corresponding to current time. $\mathbf{v}_{k, t}$ represents the velocity of the $k$-th point at time $t$, and $\hat{\mathbf{v}}_{k, t}$ represents the ground-truth.

\subsection{Basic experiment}

\begin{figure*}
    \centering
    \subfigure[]{\includegraphics[scale=0.16]{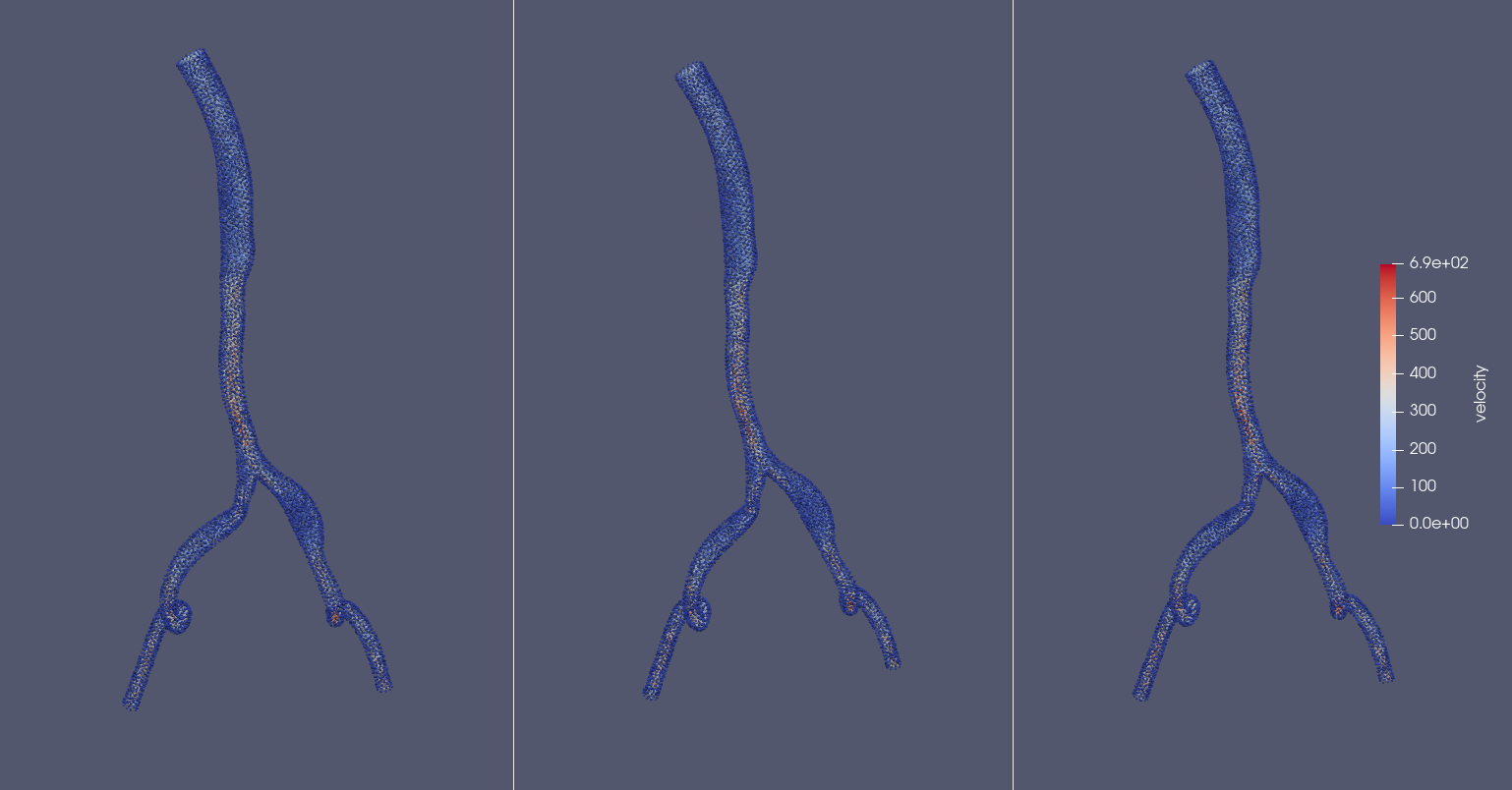}}
    \subfigure[]{\includegraphics[scale=0.16]{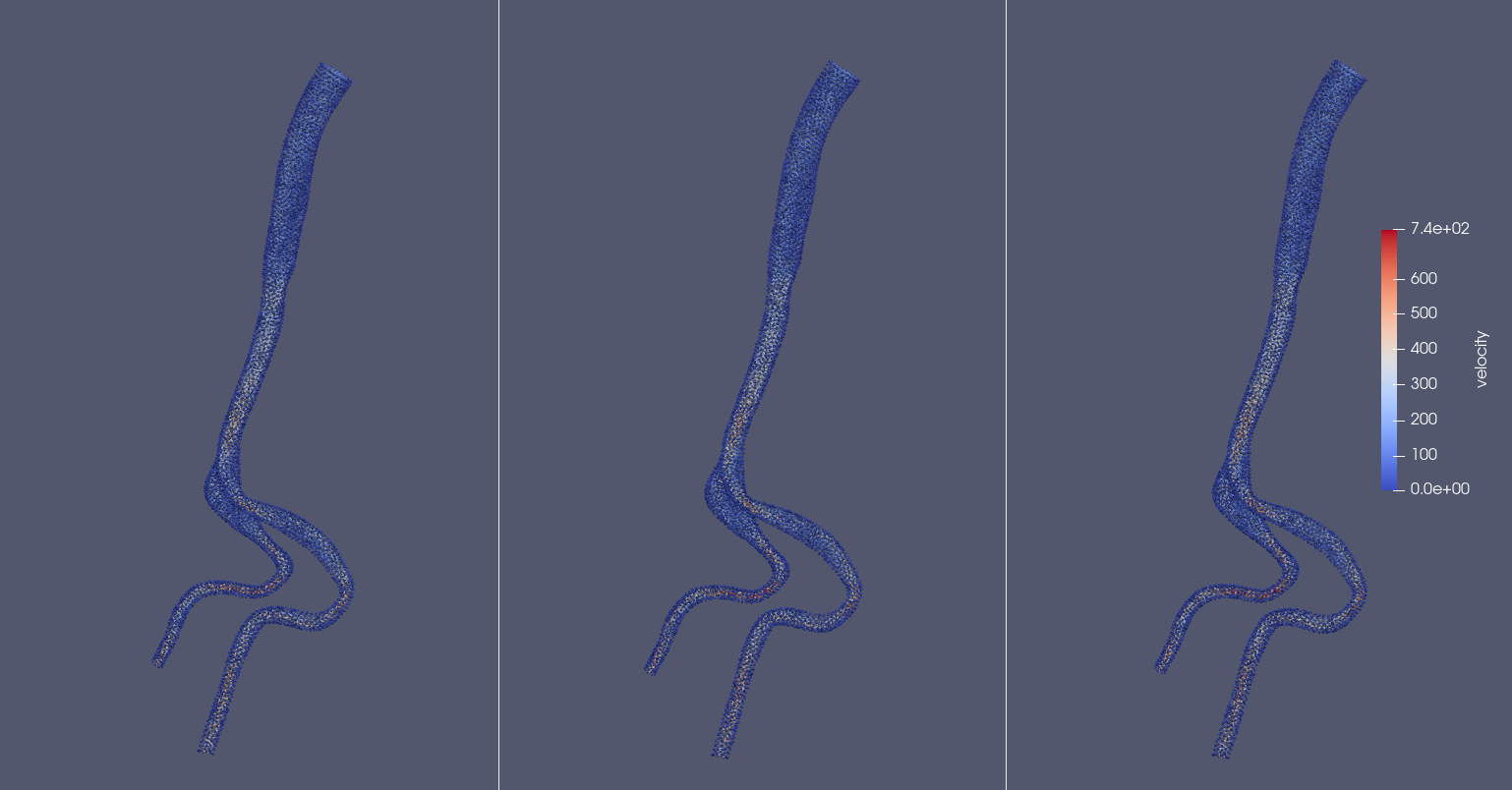}}
    \subfigure[]{\includegraphics[scale=0.3]{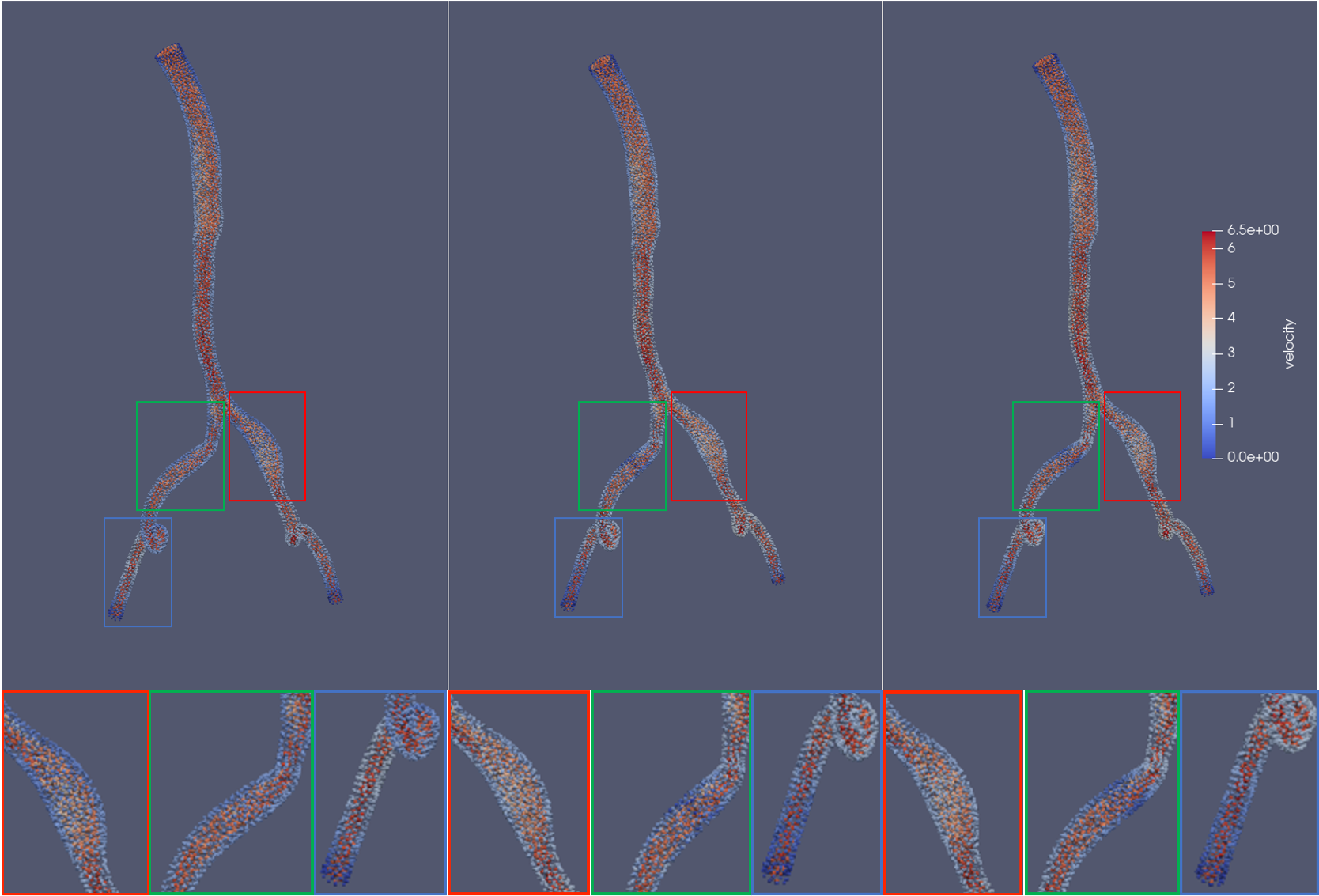}}
    \subfigure[]{\includegraphics[scale=0.3]{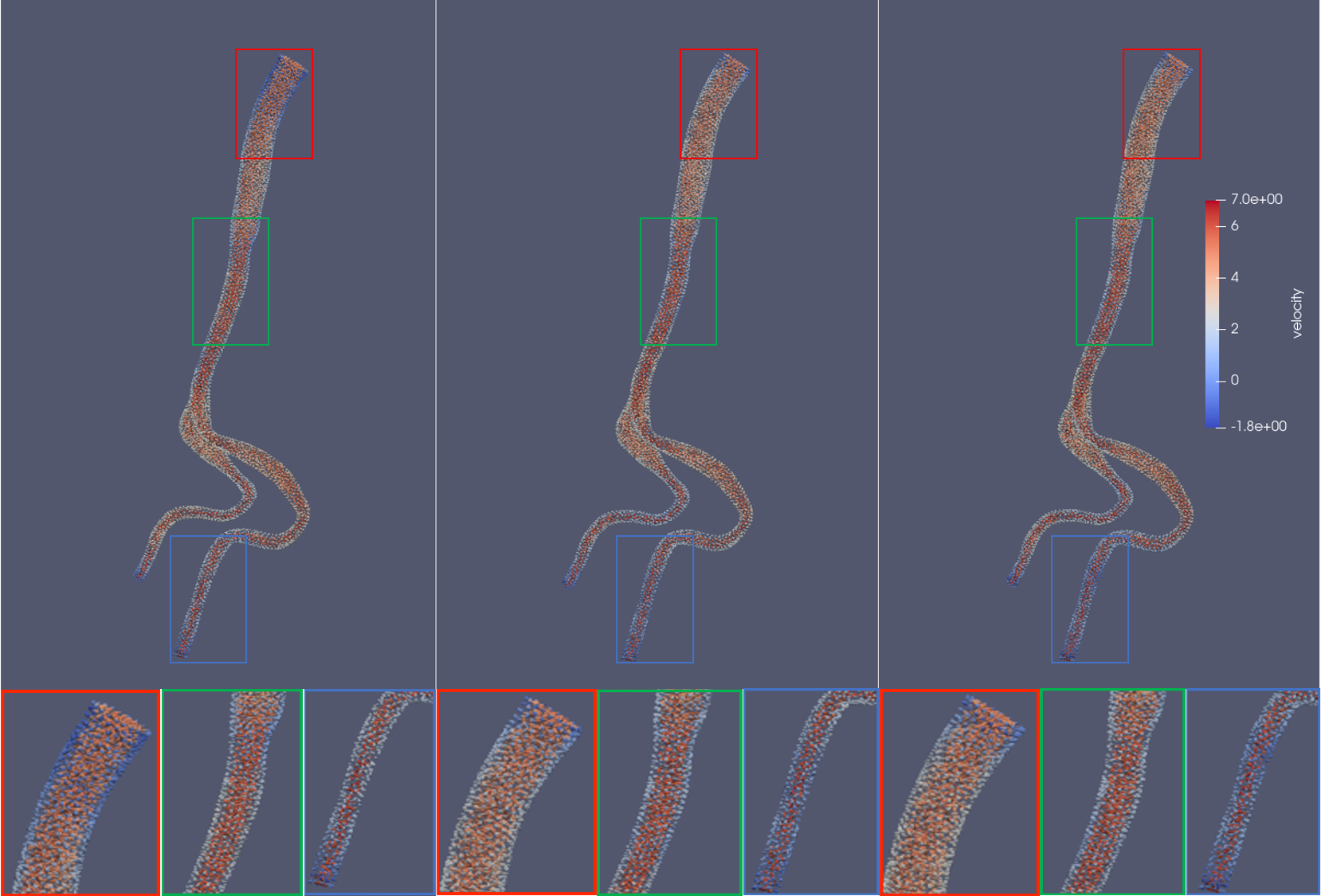}}
    \caption{Interpolation results. (a) and (b) are velocity fields at different frames generated by different resistance of the same blood vessel model. (c) and (d) are logarithmic conversion results of the velocity fields of (a) and (b) respectively, where from left to right is linear interpolation, network generation, and the ground-truth.}
    \label{fig:exp1_res}
\end{figure*}
Figure \ref{fig:exp1_res} shows the velocity fields of the same vessel with different boundary conditions and different frames. In order to compare the results of linear interpolation, network interpolation and the ground-truth in an intuitive way, we perform logarithmic conversion on the velocity field. Figure \ref{fig:exp1_res} (c) and (d) show the result of the logarithmic conversion of Figure \ref{fig:exp1_res} (a) and (b). The specially marked parts in (c) and (d) are the areas with large differences between the three. It can be clearly seen that results generated by the network are closer to the ground-truth in detail. Meanwhile, the velocity modulus length and the range of velocity components in the $x$, $y$, and $z$ directions are calculated which are shown in Table \ref{tab:vel_range}. The comparison indicates that Whether the velocity modulus length or the velocity components in each direction, the results generated by the proposed network are closer to the ground-truth.

\begin{figure}
    \centering
    \includegraphics[scale=0.52]{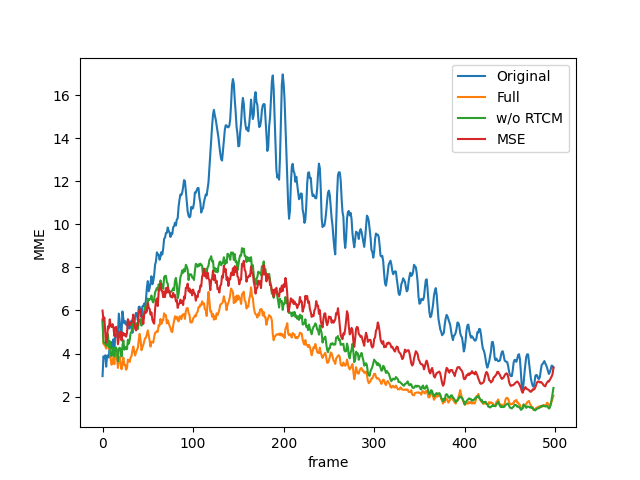}
    \caption{\emph{MME} curve. The blue curve shows error between the velocity field generated by linear interpolation and the ground-truth, the orange one is the error between the network interpolation results generated by our full model and the ground-truth, the green curve is the error between the network interpolation results generated by the model without RTCM and the ground-truth, and the red one is the error between $L_{m s e}$ training network result and the ground-truth.}
    \label{fig:exp1_mme}
\end{figure}

\begin{table}
\begin{center}
\begin{tabular}{|l|c|c|c|c|}
\hline
Modulus &Linear &Network & Ground truth \\
\hline\hline
$v$ & [0,604] & [0,675] & [0,690] \\
$v_{x}$ & [-191,313] & [-229,384] & [-220,391]  \\
$v_{y}$ & [-381,574] & [-443,607] & [-459,625] \\
$v_{z}$ & [-518,148] & [-516,228] & [-591,196] \\
\hline
\end{tabular}
\end{center}
\caption{velocity interval range. The left row is the result of linear interpolation, the mid row is the network output, and the right row is the ground-truth.}
\label{tab:vel_range}
\end{table}
For different frames of the velocity field, performance of linear interpolation and network interpolation can be shown through \emph{MME}. Figure \ref{fig:exp1_mme} shows the \emph{MME} curve of figure \ref{fig:exp1_res} (a). Except for the few frames in the beginning, the error curves generated by our network are all below the error curve of linear interpolation. From the perspective of the entire flow field shown in Figure \ref{fig:exp1_mme}, the overall error of the network interpolation is 3.715, and it is 8.574 for linear interpolation. The error of velocity field generated by the network is 56.67\% lower than the linear interpolation.

In addition, we calculate the relative error between velocity field of interpolation and the ground-truth for all generated fields. The one-frame network interpolation and linear interpolation experimental results are shown in the first and second columns of Table \ref{tab:exp1_re}. For each field, the relative error of velocity field generated by network is smaller than the linear interpolation result. Average relative error of network generation for all fields is 17.81\%, and the average relative error of linear interpolation for all fields is 34.03\%. Hence, the result generated by our network is 47.66\% less than the result of linear interpolation.

\begin{table}
\begin{center}
\begin{tabular}{|l|c|c|c|c|}
\hline
Field & Full & Linear & w/o RTCM & $L_{MSE}$ \\
\hline\hline
Aorta1+R2000 & \textbf{17.26} & 26.67 & 17.97 & 30.21 \\
Aorta1+R8000 & \textbf{16.25} & 36.50 & 16.61 & 30.41 \\
Aorta2+R2500 & \textbf{19.60} & 39.64 & 20.39 & 34.22 \\
Aorta2+R3000 & \textbf{18.03} & 38.02 & 18.83 & 32.70 \\
Aorta3+R1500 & 19.09 & 26.03 & \textbf{18.98} & 34.06 \\
Aorta3+R6500 & \textbf{14.58} & 30.40 & 14.78 & 29.50 \\
Aorta4+R1500 & \textbf{18.18} & 29.44 & 18.24 & 29.51 \\
Aorta4+R7000 & 14.79 & 39.90 & \textbf{14.65} & 26.02 \\
Aorta5+R1500 & \textbf{24.40} & 41.55 & 24.64 & 40.42 \\
Aorta5+R4000 & 15.92 & 32.14 & \textbf{15.89} & 33.56 \\
Average & \textbf{17.81} & 34.03 & 18.10 & 32.06   \\
\hline
\end{tabular}
\end{center}
\caption{Relative errors of different blood vessel flow samples. Here, \emph{Full} denotes our full model, \emph{Linear} denotes the linearly interpreted high-resolution flows, \emph{w/o RTCM} denotes our model variant without the resistance-time encoder, $L_{MSE}$ denotes our model trained with the MSE loss instead of the magnitude loss and orientation loss.}
\label{tab:exp1_re}
\end{table}


\subsection{Effect of Resistance-Time Co-modulation}
\label{ssec:exp_rscm}
In order to verify the effect of the Resistance-Time Co-modulation module , we conduct comparative experiments. Specially, we removed the resistance-time encoder from the full model, then train and test it under the same settings. We compare the degraded model with the full model in terms of the average modulus length error and relative error.

The single frame \emph{MME} of the flow field is shown in Figure \ref{fig:exp1_mme}, where the orange curve is the \emph{MME} error w.r.t. our full model, and the green curve is the MME error w.r.t. the model without RTCM. Among them, the initial (blue) average error is 8.574, the average error of the flow field without RTCM is 4.549, and the average error of the flow field with RTCM is 3.715. From the perspective of the average error of the velocity field, the result of adding RTCM is 18.33\% lower than that of not adding it. From the curve point of view, after RTCM is added, the average modulus length error of each frame of the velocity field is smaller than the result without RTCM, and much smaller than the linear interpolation result. this also proves the effectiveness of RTCM.

The results of velocity field \emph{RE} are shown in columns 1, 2 and 3 in Table \ref{tab:exp1_re}. For the \emph{RE} of each velocity field in the generated data, the \emph{RE} of adding RTCM is mostly smaller than the result of not adding RTCM. From the overall generated data, the relative error of the network with RTCM is 17.81\%, and the relative error without RTCM is 18.10\%, a relative decrease of 1.60\%.

\subsection{Loss verification}
\label{ssec:exp_loss}
In order to prove that $L_{mo}$ is more in line with our task, we compare it with MSE loss. We can train the proposed DL model through different Loss, and then compare the generated results through \emph{MME} and \emph{RE}. 

The comparison result of \emph{MME} is shown in Figure \ref{fig:exp1_mme}. The red curve is the single-frame \emph{MME} generated by $L_{m s e}$ training, and the orange curve is the single-frame \emph{MME} generated by $L_{mo}$ training. It can be found from the figure that the \emph{MME} curve of a single frame generated by $L_{mo}$ training is below the curve generated by $L_{m s e}$ training, and the \emph{MME} at any frames is less than the result generated by $L_{m s e}$ training. We calculate the overall average \emph{MME} of the velocity field, the initial (blue) average error is 8.574, the average error of the velocity field generated by $L_{m s e}$ is 5.086, and the average error of the velocity field generated by $L_{mo}$ training is 3.715. Compared with the results generated by $L_{m s e}$ training, there is a relative decrease of 26.96\%. 

We can also compare the performance of the two on the overall generated data. The \emph{RE} results are shown in Column 1, 2, and 4 in Table \ref{tab:exp1_re}. It can be seen from the table that the relative errors of the results generated by $L_{mo}$ training on all generated data are much smaller than the results generated by $L_{m s e}$. The overall relative error of the results generated by $L_{m s e}$ training on the generated data is 32.06\%, and the overall relative error of the results generated by  $L_{mo}$ training on the generated data is 17.81\%, a relative decrease of 44.45\%. From this, we can see that $L_{mo}$ is more suitable for the current task.

\subsection{Two-frame Interpolation}
\label{ssec:exp_2frame}

\begin{figure*}
    \centering
    \subfigure[]{\includegraphics[scale=0.5]{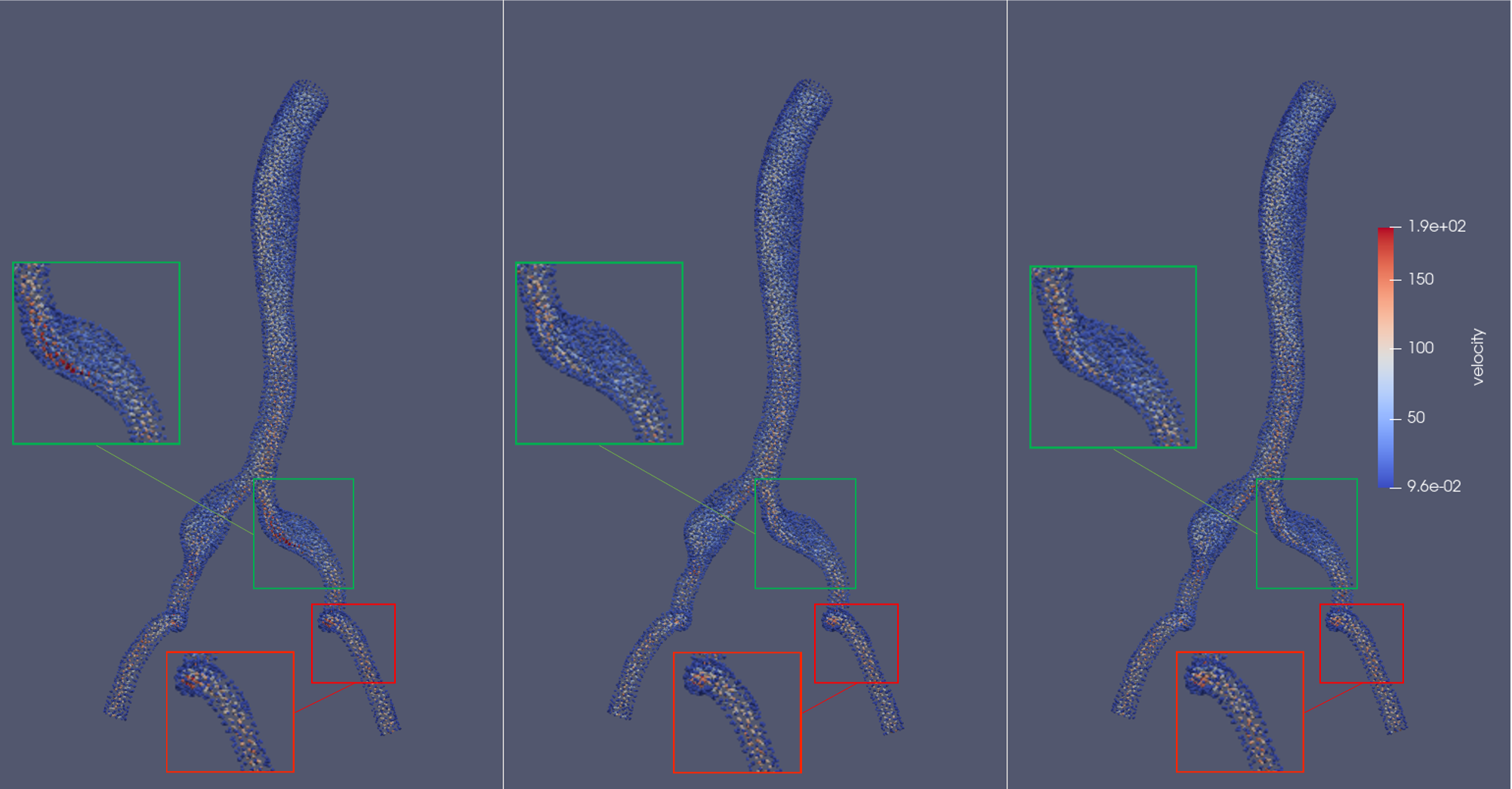}}
    \subfigure[]{\includegraphics[scale=0.5]{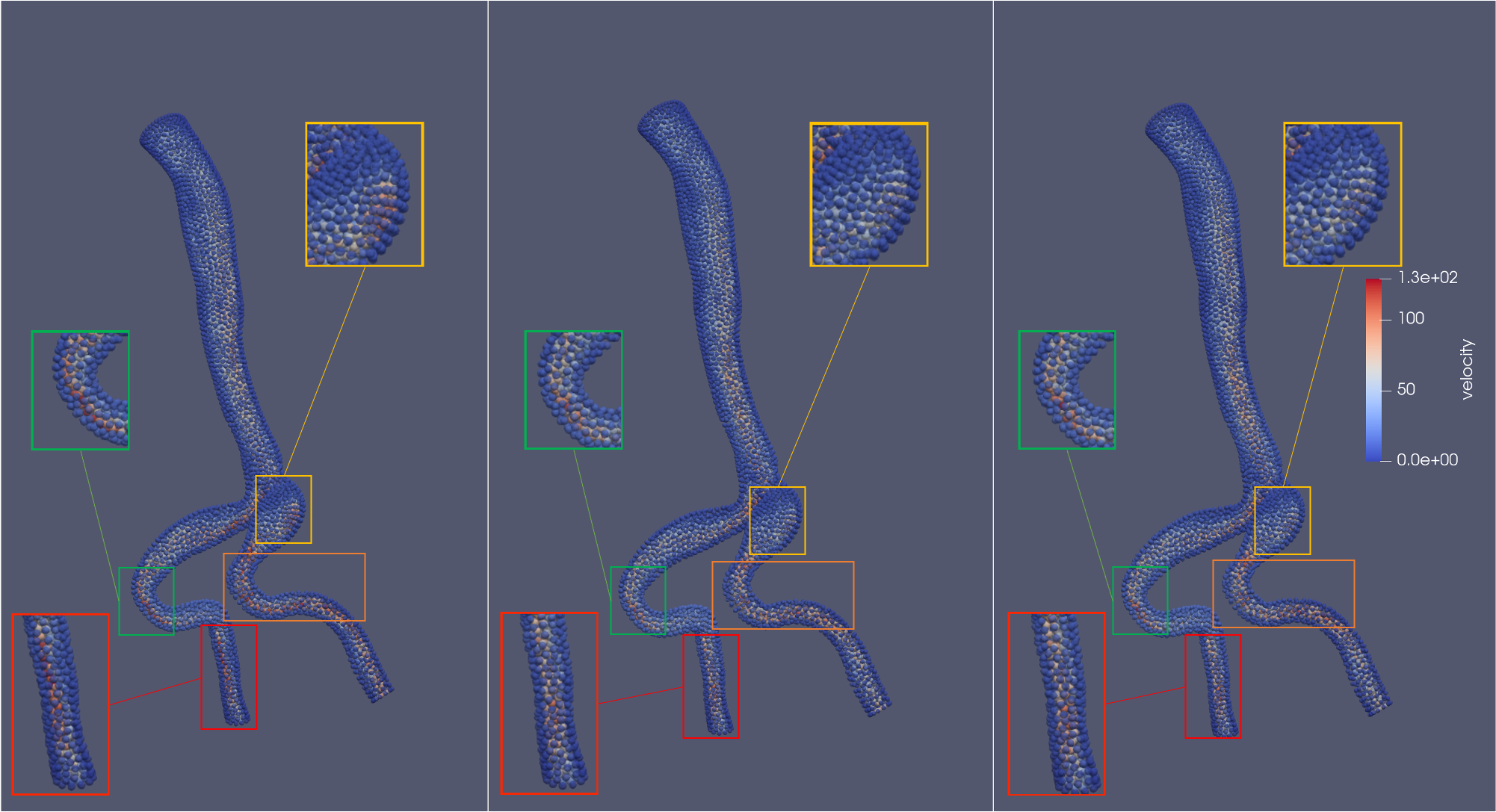}}
    \caption{Two-frame interpolation results. (a) and (b) are the simulation results of one frame of velocity field under different blood vessel models. From left to right: the linear interpolation result, the network generation result and the ground-truth.}
    \label{fig:exp2_res}
\end{figure*}

\begin{figure}
    \centering
    \includegraphics[scale=0.55]{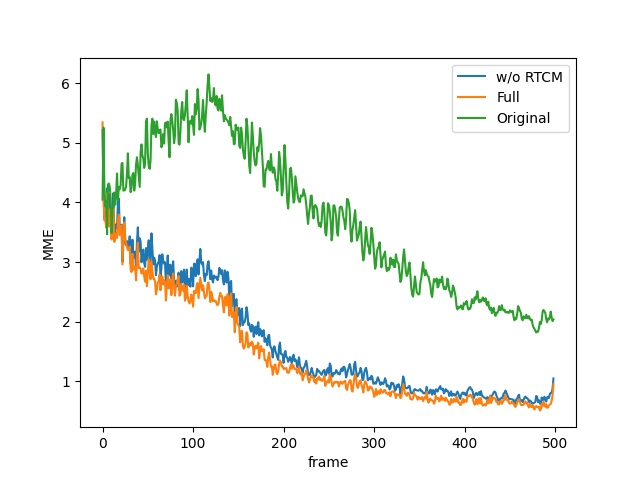}
    \caption{\emph{MME} curve. The green curve shows error between the velocity filed generated by linear interpolation and the ground-truth, the blue one is the error between the result generated by our model without RTCM and the ground-truth, and the orange one is the error between the result generated by our full model and the ground-truth.}
    \label{fig:exp2_mme}
\end{figure}

In addition to the basic experiment of one-frame interpolation, we also perform two-frame interpolation using a low-accuracy velocity field as the network input to reconstruct its corresponding high-accuracy, high-temporal-resolution velocity field. The input of the network is $\left[\mathbf{u}_{i, t}, \mathbf{u}_{i, t+1}, \mathbf{l}_{i}\right] \in \mathbb{R}^{3 \times 3}$, and the network output is $\left[\mathbf{v}_{i, t}, \mathbf{v}_{i, t+1 / 3}, \mathbf{v}_{i, t+2 / 3}, \mathbf{v}_{i, t+1}\right] \in \mathbb{R}^{4 \times 3}$.

Figure \ref{fig:exp2_res} shows the visualization results of interpolating two frames, where (a) and (b) are the velocity fields of a certain frame for different blood vessel models. The left column is the linear interpolation result, the middle column is the network generation result, and the right column is the high-accuracy ground truth. Figure \ref{fig:exp2_res} clearly shows that in terms of local details, the results generated by the network are closer to ground truth than the results using linear interpolation.

\begin{table}
\begin{center}
\begin{tabular}{|l|c|c|c|c|}
\hline 
Modulus &Linear &Network & Ground truth \\
\hline\hline
$v$ & [0,129] & [0,118] & [0,125] \\
$v_{x}$ & [-67,80] & [-61,82] & [-59,72]  \\
$v_{y}$ & [-60,120] & [-82,113] & [-79,114] \\
$v_{z}$ & [-116,33] & [-103,25] & [-105,27] \\
\hline
\end{tabular}
\end{center}
\caption{Velocity interval range. The left row is the result of linear interpolation, the mid row is the network output, and the right row is the ground-truth.}
\label{tab:exp2_range}
\end{table}

In addition, we calculate the modulus length of the velocity field shown in Figure \ref{fig:exp2_res} (b) and the range of velocity components in the $x$, $y$, and $z$ directions, as shown in Table \ref{tab:exp2_range}. The velocity field output by the network is closer to ground truth than the traditional linear interpolation method in terms of modulus length and various direction components.

Figure \ref{fig:exp2_mme} shows the \emph{MME} curve of two-frame interpolation experiment. The green curve is the initial error, where the average \emph{MME} of the velocity field is 3.734. The blue curve is the velocity field error generated by the model without RTCM, where the average \emph{MME} of the velocity field is 1.654. The orange curve is the velocity field error generated by our full model, and the average \emph{MME} is 1.466. It can be seen from the figure that compared to linear interpolation, the results generated by our network demonstrate an accuracy improvement, and the accuracy is improved by 60.74\%. In addition, comparing the velocity field results with or without the addition of RTCM, there are obvious differences in the \emph{MME} curve. The single frame error of adding RTCM is less than that of not adding RTCM, which further proves that the RTCM has a positive effect on the velocity field error reduction.

At the same time, we analyze the \emph{RE} results of two-frame interpolation, as shown in Table \ref{tab:exp2_re}. For the entire generated data, the \emph{RE} of the naive linear interpolation method is 33.61\%, while the \emph{RE} without RTCM is 21.17\%, which is 12.44\% lower than that of the linear interpolation method, and the \emph{RE} with RTCM is 20.63\%, compare to the \emph{RE} of the linear interpolation method is reduced by 12.98\%. From the generated data of each velocity field, the network generated results (the right and middle columns in Table \ref{tab:exp2_re}) have a smaller \emph{RE} than linear interpolation. The \emph{RE} of adding RTCM is even smaller than that of not adding RTCM.

\begin{table}
\begin{center}
\begin{tabular}{|l|c|c|c|}
\hline
Field & Full & Linear &  w/o RTCM \\
\hline\hline
Aorta1+R2000 & \textbf{19.00} & 26.30 & 20.28  \\
Aorta1+R8000 & \textbf{18.33} & 36.01 & 19.34  \\
Aorta2+R2500 & \textbf{22.82} & 39.29 & 22.95  \\
Aorta2+R3000 & 21.32 & 37.63 & \textbf{21.23}  \\
Aorta3+R1500 & \textbf{21.19} & 25.87 & 21.35  \\
Aorta3+R6500 & \textbf{16.93} & 30.11 & 19.20  \\
Aorta4+R1500 & 21.45 & 28.95 & \textbf{21.00}  \\
Aorta4+R7000 & 17.86 & 39.10 & \textbf{17.75}  \\
Aorta5+R1500 & \textbf{27.96} & 40.99 & 28.86  \\
Aorta5+R4000 & \textbf{19.39} & 31.81 & 19.76  \\
average & \textbf{20.63} & 33.61 & 21.17 \\

\hline
\end{tabular}
\end{center}
\caption{Relative Error}
\label{tab:exp2_re}
\end{table}

\subsection{Advantage in Time Consumption}
In the data set production process, SimVascular~\cite{lan2018re} is used for simulation, the CPU is Intel Core i5, the simulation process is parallelized through MPI, and the number of processes is set to 8. The time-stepping scheme used in the simulation is the generalized alpha time-stepping scheme~\cite{jansen2000generalized}. Under the same conditions, it takes about 2 hours and 07 minutes to construct a 500-frame high-accuracy time-varying velocity field. It takes about 26 minutes to construct a 250-frame low-accuracy time-varying velocity field.

Our network runs on a server equipped with Intel Xeon CPU E5-2620 and Titan X Pascal GPU, and it takes about 21 seconds to generate a 500-frame time-varying velocity field on average.

\begin{figure}
    \centering
    \includegraphics[scale=0.55]{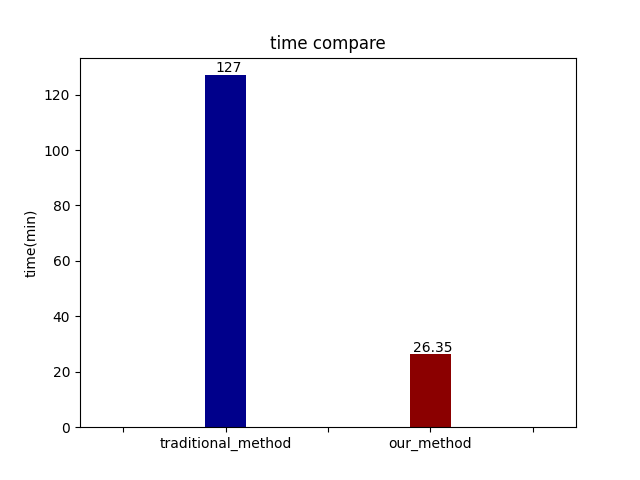}
    \caption{Comparison of time consumption.}
    \label{fig:time}
\end{figure}

Therefore, using the traditional method to generate a high-accuracy time-varying velocity field (500 frames) takes 127 minutes, while using our method to generate a high-accuracy time-varying velocity field (500 frames) only takes 26.35 minutes. The time consumption is compared in Figure \ref{fig:time}, and the simulation speed is increased by nearly 5 times.


\section{Conclusion}
\label{sec:conclude}

In this paper, a novel deep learning framework is proposed for temporal super-resolution simulation of blood vessel flows, in which a high-temporal-resolution time-varying blood vessel flow simulation is generated from a low-temporal-resolution  flow simulation result. In our framework, point-cloud is used to represent the complex blood vessel model, resistance-time aided PointNet model is proposed for extracting the time-space features of the time-varying flow field, and finally we can reconstruct the high-accuracy and high-resolution flow field through the Decoder module. The corresponding high-resolution flow fields were reconstructed with high accuracy.  Several examples are given to illustrate the effective and efficiency of the proposed framework for temporal super-resolution simulation of blood vessel flows.  

Overall, this work may contribute to a robust and very general framework for the development of the blood simulation model in patient-specific medical applications and enrich the application of super-resolution technology in fluid mechanics.

{\small
\bibliographystyle{cvm}
\bibliography{cvmbib}
}

\end{document}